\documentclass[10pt,a4paper,twocolumn,multicol,superscriptaddress,prb,amsmath,amssymb,aps,prb]{revtex4}
\usepackage{graphicx}
\usepackage{bm}

\begin{document}
\title{Charge and Spin Hall Conductivity in Metallic Graphene.} 


\author{N.A. Sinitsyn}
\affiliation{Department of Physics, University of Texas at Austin,
Austin TX 78712-1081, USA}
\author{J.E. Hill} 
\affiliation{Department of Physics, University of Texas at Austin,
Austin TX 78712-1081, USA}
\author{Hongki Min} 
\affiliation{Department of Physics, University of Texas at Austin,
Austin TX 78712-1081, USA}
\author{Jairo Sinova}
\affiliation{Department of Physics, Texas A\&M University,
College Station, TX 77843-4242, USA}
\author{A.H. MacDonald} 
\affiliation{Department of Physics, University of Texas at Austin,
Austin TX 78712-1081, USA}

\date{\today}

\begin{abstract}
Graphene has an unusual low-energy band structure with four chiral bands and
half-quantized and quantized Hall effects that have recently attracted theoretical
and experimental attention.  We study the Fermi energy and disorder dependence
of its spin Hall conductivity $\sigma^{SH}_{xy}$.  In the metallic regime we find that 
vertex corrections enhance the intrinsic spin Hall conductivity and that 
skew scattering can lead to $\sigma^{SH}_{xy}$ values that exceed the quantized ones 
expected when the chemical potential is inside the spin-orbit induced energy gap.
We predict that large spin Hall conductivities
will be observable in graphene even when the spin-orbit gap does not survive disorder.  
\end{abstract}
\maketitle

\noindent
{\em Introduction}-- 
The low-energy band structure of graphene consists of four chiral bands that 
realize (2+1)-dimensional relativistic field theory models with 
parity anomalies.  The anomalies imply unusual spectra in an external magnetic field
and quantized and half-quantized Hall effects\cite{{Semenoff84},{Haldane87}}. 
Theoretical interest\cite{graphene_old} in these unusual electronic systems has increased\cite{graphene_new} recently 
because of experimental progress\cite{graphene_exp}, including measurements of the anticipated {\em half-quantized} 
quantum Hall effect. One particularly interesting observation, due to Kane and Mele \cite{{KaneQSHE},{KaneZ2}}, 
is that because of a gap produced by spin-orbit interactions, the spin Hall conductivity $\sigma_{SH}$ of undoped graphene
is quantized {\em in the absence of a magnetic field}. This suggestion is related to recent work
on the anomalous Hall effect in ferromagnetic metals\cite{ferrointrinsicahe} and on its paramagnetic cousin, 
the spin Hall effect\cite{intrinsicshe}, in which it was
suggested that these 
transport coefficients can be dominated by 
an intrinsic momentum-space Berry phase contribution that 
reduces to quantized values when the Fermi level is in a gap.  Here we examine how the quantized spin
Hall effect is altered when the Fermi energy in the graphene plane is gated into
the metallic regime.  We find that the intrinsic spin Hall effect is no longer quantized, that 
it is enhanced by disorder vertex corrections, and that in the metallic regime skew scattering 
can potentially lead to parametrically larger spin Hall conductivities.  Because the 
Bloch state disorder broadening in current samples is (according to our estimates) much larger 
than the clean system spin-orbit gap, these results are necessary for the 
interpretation of experiment.  Spin-Hall effects should be observable even when the 
spin-orbit gap does not survive disorder.  

\noindent
{\em Disordered Graphene Model}-- When spin-orbit interactions are included,\cite{KaneQSHE} the low-energy 
physics of a clean undoped graphene crystal is described by an eight-band envelope function Hamiltonian 
\begin{equation}
\hat{H_0}=v(k_x \tau_z\sigma_x +k_y \sigma_y) +\Delta \sigma_z \tau_z s_z
\label{dh1c}
\end{equation}
where $s_z=\pm$ is the up/down electron spin component perpendicular to the graphene plane, 
$\tau_z=\pm$ is a valley label that specifies one of the two inequivalent ($K$ and $K'$) points in the 
crystal Brillouin zone near which low-energy states occur, and the $\sigma_i$ are
Pauli matrices representing a pseudo-spin degree of freedom corresponding to the 
two sites per primitive cell of a hexagonal lattice. The parameter $\Delta$ is the strength of the spin-orbit
coupling and we take $\hbar=1$.  For $\Delta = 0$ this Hamiltonian defines four 
spin-degenerate gapless bands in which the pseudospin orientation lies in the ${\hat x}$-${\hat y}$ plane and 
winds around the $\hat z$-axis, either clockwise or counter-clockwise, with a $2 \pi$ planar wavevector rotation.
The operators $\sigma_i$, $s_z$ and 
$\tau_z$ commute with each other. Random defects can in general produce transitions between bands and 
between spins.  Here we assume spatially smooth spin-independent disorder so that $s_z$ and 
$\tau_z$ are good quantum numbers, allowing us to consider the cases $\tau_z, s_z = \pm 1$ independently.
For this disorder model
we evaluate the Kubo-formula Hall conductivity in the self-consistent 
Born approximation (SCBA) for chemical potentials inside and outside the spin-orbit gap, 
including both non-trivial pseudospin dependent disorder self-energies and ladder 
diagram vertex corrections. When the chemical potential lies in the gap,
an elementary calculation shows that in the absence of disorder the single-band bulk partial Hall
conductivity is given exactly by the half-quantized Berry phase
contribution,\cite{{ferrointrinsicahe},{intrinsicshe}} $-(s_ze^2/{2h}) $.
disorder corrections to intrinsic Hall effect are small near the gap edge but yield substantial enhancement 
in more strongly gated systems.

\noindent
{\em $2D$-Dirac-band Hall effect}--The 2D Dirac Hamiltonian in the spin $\uparrow$ 
$K$-valley is
\begin{equation}
\hat{H}= v(k_x \sigma_x +k_y \sigma_y) +\Delta \sigma_z,
\label{dh1}
\end{equation}
Spin-orbit-coupling opens up a gap which breaks the spectrum into an electron band at positive energies and a 
hole band at negative energies  
$\epsilon_{{\bf k}}^{\pm}=\pm \sqrt{\Delta^2+(v k)^2}$,
where $k=|{\bf k}|$  and $\pm$ refer to electron and hole bands respectively.
(The three other graphene bands differ either in the Dirac band chirality sense, or in
the sign of the mass term, or in both ways.) 
In what follows, we assume that the Fermi energy is positive; because of the symmetry of the Dirac Hamiltonian
generalization to negative $\epsilon_F$ is trivial. 

The Kubo formula for the Hall conductivity depends on both band-diagonal and off-diagonal matrix elements of the 
velocity operator and on the electronic Green's function.  The disorder-free retarded Green's function and velocity operators for
this Hamiltonian are  $G^R_0(\epsilon) =(\epsilon-\hat{H}+i\eta)^{-1}$,
$v_x=v \sigma_x$, and $v_y=v \sigma_y$.  
It will prove convenient to use the Streda-Smrcka\cite{Streda} version of the Kubo
formula which separates Fermi surface and occupied state contributions:
$\sigma_{xy}=\sigma^{I}_{xy}+\sigma^{II}_{xy}$
where
\begin{eqnarray}
\noindent \sigma_{xy}^{I}&=&\frac{-e^2}{4\pi} \int_{-\infty}^{+\infty} d\epsilon \frac{df(\epsilon)}{d\epsilon} \mathrm{Tr}[v_x (G^R(\epsilon)-
G^A(\epsilon))v_yG^A(\epsilon)\nonumber\\
&& -v_x G^R(\epsilon) v_y (G^R(\epsilon)-G^A(\epsilon))]
\label{sigmaI}
\end{eqnarray}
and
\begin{eqnarray}
\noindent \sigma_{xy}^{II}=\frac{e^2}{4\pi} \int_{-\infty}^{+\infty} d\epsilon f(\epsilon) \mathrm{Tr}[v_x G^R(\epsilon) v_y \frac{G^R(\epsilon)}{d\epsilon} -v_x \frac{G^R(\epsilon)}{d\epsilon}\nonumber\\ 
\times v_yG^R(\epsilon)
-v_x G^A(\epsilon) v_y \frac{G^A(\epsilon)}{d\epsilon} 
+v_x\frac{G^A(\epsilon)}{d\epsilon}v_yG^A(\epsilon)]
\label{sigmaII}
\end{eqnarray}
\noindent
{\em 2D-Dirac band intrinsic Hall conductivity}---
The Hall conductivity in the absence of disorder is most simply evaluated 
by expressing\cite{ferrointrinsicahe} it in terms of matrix elements of velocity operator
between unperturbed Bloch states: 
\begin{equation}
\begin{array}{l}
\sigma_{xy}^{int}=\frac{e^2}{\Omega}\sum_{{\bf k}} \frac{f_{{\bf k}}^{+} - f_{{\bf k}}^{-}}{(\epsilon_{{\bf k}}^+ - \epsilon_{{\bf k}}^-)^2} 
2{\rm Im} [ \langle  u_{{\bf k}}^{-} | v_y | u_{{\bf k}}^{+} \rangle \langle  u_{{\bf k}}^{+} | v_x|  u_{{\bf k}}^{-} \rangle ]
\end{array}
\label{sigint}
\end{equation}
where the $f_{\bf k}^{\pm}$ are occupation numbers in the electron and hole bands, 
$\Omega$ the area of the system, and $|u^{\pm}_{{\bf k}} \rangle$
the ${\bf k}$-dependent pseudospinors of the chiral Dirac Hamiltonian, Eq.(~\ref{dh1}).
\begin{equation} 
|u^{+}_{{\bf k}}\rangle=\left( \begin{array}{l}
cos(\theta/2) \\
sin(\theta /2) e^{i\phi} 
\end{array} \right),\,\,\,\, 
|u^{-}_{{\bf k}}\rangle=\left( \begin{array}{l}
sin(\theta/2) \\
-cos(\theta /2) e^{i\phi} 
\end{array} \right)
\label{basis}
\end{equation}
where  $\cos(\theta)=\Delta/\sqrt{(vk)^2+\Delta^2}$,  
and $\tan(\phi) = k_y/k_x$.
For the chemical potential in the upper band with Fermi momentum $k_F$ we find
\begin{equation}
\sigma_{xy}^{int} = -\frac{e^2\Delta }{4\pi \sqrt{(v k_F)^2+\Delta^2}} \, ,
\label{sigint1}
\end{equation}
Eq.(\ref{sigint1}) includes $\sigma_{xy}^{II}$ and
the disorder free limit of $\sigma_{xy}^{I}$.
In the metallic regime the disorder-independent part of $\sigma^{I}_{xy}$ equals with (\ref{sigint1}) 
so that in this regime $\sigma^{II}_{xy}=0$.

When the chemical potential is in the gap ($k_F \to 0$) 
\begin{equation}
\sigma_{xy} \to \sigma^{gap}_{xy} \equiv -\frac{e^2}{4\pi}.
\label{sigmaIIfin}
\end{equation}
This is the 2D-Dirac model's half quantized (in units $e^2/2\pi\hbar$) Hall conductivity, which after summing over bands 
is responsible for the quantum spin-Hall-effect discussed in Refs.[\onlinecite{{KaneQSHE},{KaneZ2},{Haldane_gr_num}}]. 
It should seem surprising
that the Hall conductivity (\ref{sigmaIIfin}) is only half-integer given general arguments that the 
Hall conductance of non-interacting electrons must be quantized.
The resolution of this paradox is that bands come in pairs.  The sum of the K and K' valley bulk conductivities is quantized;
correspondingly only one band of edge states is induced by the truncation 
of both K and K' bulk bands.
  
\noindent
{\em Influence of Disorder on $\sigma_{xy}$:}---
We assume a $\delta$-correlated spin-independent random potential with Gaussian correlations
 $\langle V({\bf r}_1) V({\bf r}_2) \rangle_{dis} =nV_0^2 \delta ({\bf r}_1 -{\bf r}_2)$.  
\begin{figure}[h]
\includegraphics[width=4cm]{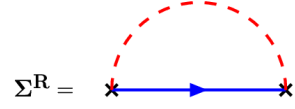}
\centering
\caption{Self-energy Feynman diagram in self-consistent Born approximation.}
\label{selfen1}
\end{figure}
The SCBA that we employ includes only contributions from Feynman diagrams
without crossed disorder correlation lines.  This common approximation is self-consistent
but is 
incomplete.  We assume that crossed-disorder-line contributions give rise to parametrically distinguishable
effects and do not affect our qualitative conclusions
about Hall effects in metallic graphene.
Fig.\ref{selfen1} illustrates the SCBA self-energy diagram which can be evaluated to obtain  
$\Sigma^{R}=-\frac{i}{4\tau^q} (1+\sigma_z \cos(\theta))$
where $\tau^q$ is a quantum life time at the Fermi surface:
\begin{equation}
1/\tau^q = nV_0^2 \int k \, dk  \, \delta(\epsilon_F-\epsilon^+_{{\bf k}})=\frac{nV_0^2k_F}{v}.
\label{tauq}
\end{equation}
Following the notation of Dugaev {\em et al.} \cite{Dugaev}, 
the SCBA retarded Green's function is
\begin{equation}
\begin{array}{l}
G^R=\frac{1}{1/G_0^R - \Sigma^{R}}
=\frac{\epsilon_F+i\Gamma_0 +v(k_x \sigma_x +k_y \sigma_y)+(\Delta - i \Gamma_1) \sigma_z}
{(\epsilon_F-\epsilon^+ +i\gamma^+)(\epsilon_F -\epsilon^- +i\gamma^-)}
\end{array}
\label{gf2}
\end{equation}
where $\Gamma_0=1/(4\tau^q)$, $\Gamma_1 = \Gamma_0 \cos(\theta)$, $\gamma^+ =\Gamma_0 +\Gamma_1 \cos(\theta)$, $\gamma^-
 =\Gamma_0 -\Gamma_1 \cos(\theta)$.  For these chiral bands disorder not only gives the 
 quasiparticle states a finite lifetime but also changes the quasiparticle eigenspinors.  
\begin{figure}[h]
\includegraphics[width=7cm]{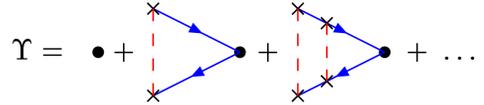}
\centering
\caption{Vertex correction Feynman diagram.  Black dots represent the Pauli operator. }
\label{fvertex}
\end{figure}
The SCBA for two-particle Green's functions like the Hall conductivity includes in addition ladder diagram vertex correction
illustrated in Fig.\ref{fvertex}. For large $v k_F \tau$ the terms in $\sigma_{xy}^{I}$ which are products of retarded and advanced Green's 
functions dominate so that the 2D matrix vertex function for which we must solve satisfies: 
\begin{equation}
\Upsilon_y=\sigma_y+nV_0^2 \int \frac{d^2 {\bf k}}{(2\pi)^2} \, G^R \Upsilon_y G^A.
\label{ve}
\end{equation} 
This equation is most easily solved by assuming that  
\begin{equation}
\Upsilon_y=a\sigma_0 +b\sigma_x +c \sigma_y +d \sigma_z
\label{v1}
\end{equation}
and deriving equations for $a$, $b$, $c$, and $d$. 
We find that
\begin{equation}
\begin{array}{l}
c=\frac{2((vk)^2+2\Delta^2)}{4 \Delta^2 + (vk)^2},\,\,\,\,\,
b=-\frac{8\Gamma_0 \Delta ((vk)^2+2\Delta^2)}{(4\Delta^2 +(vk)^2)^2},
\end{array}
\label{ab}
\end{equation}
and that $a = d = 0$. 
The SCBA $\sigma_{xy}^{I}$ is obtained by substituting the disorder-dressed Green's function (Eq.(\ref{gf2})) for the bare Green's
function and $v\Upsilon_y$ for $v_y$ in the Kubo formula Eq.(\ref{sigmaI}).  We find that 
\begin{equation}
\begin{array}{l}
\sigma_{xy}=\frac{-e^2\Delta }{4\pi \sqrt{(vk_F)^2+\Delta^2}} [1+ \frac{4(vk_F)^2}{4\Delta^2+(vk_F)^2}+
\frac{3(vk_F)^4}{(4\Delta^2+(vk_F)^2)^2} ].
\end{array}
\label{sxy2}
\end{equation} 
The second and third terms in square brackets in Eq.(~\ref{sxy2}) represent  
disorder corrections to the intrinsic Hall conductivity of the 2D-Dirac model.
We note that all terms are independent of the disorder potential strength and of the concentration of 
scatterers and in this sense are parametrically similar.
They do however have different dependences on the position of the Fermi level. 
Note that when the chemical potential approaches the gap the intrinsic 
contribution remains finite and disorder corrections vanish, recovering the 
model's half quantized Hall effect. 

\noindent 
{\em Non-Gaussian disorder.}---
We have so far made the usual approximation of assuming Gaussian disorder correlations. 
Although normally small, non-zero third moments of the disorder potential distribution,
can\cite{Luttinger,Smit56} alter $\sigma_{xy}$ qualitatively since they can favor 
scattering with a particular chirality (skew scattering) and consequently lead to a $\sigma_{xy}$ contribution
that diverges in the limit of weak disorder scattering.  The size of this contribution to $\sigma_{xy}$ 
is particularly difficult to estimate since it depends very strongly on the details of the scattering 
potential.  To illustrate its potential role we consider for concreteness a model of uncorrelated $\delta$-function
scatterers: $V({\bf r}) =\sum_i V_i \delta ({\bf r}-{\bf R}_i)$, 
$R_i$ random, $\langle V_i \rangle =0$, $\langle (V_i)^2 \rangle =V_0^2 \ne 0$ 
and $\langle (V_i)^3 \rangle = V_1^3 \ne 0$. 

Given asymmetric scattering, skew scattering is a more physically transparent contribution to the 
Hall conductivity and can be described directly using either Boltzmann transport theory or the Kubo
formula, including the non-standard Feynman diagrams implied by non-Gaussian disorder models.
We apply results which have been derived previously to the graphene case. 
Let $\psi^{+}_{{\bf k}} =(1/\sqrt{\Omega}) e^{i{\bf kr}} |u^{+}_{{\bf k}}\rangle$ be a Bloch state
in the electron band with positive energy and $V_{{\bf k,k'}} =
\langle \psi^{+}_{{\bf k}}| \hat{V} |\psi^{+}_{{\bf k'}} \rangle$ be a disorder potential matrix elements within the band. Then, 
following Eqs.(32)-(36) in Ref.[\onlinecite{Loss}] for the case of zero temperature and a single band we find that 
\begin{equation}
 \frac{\sigma_{xy}^{sk}}{(e \tau^{tr})^2}=-\int \frac{d^2{\bf k}}{(2\pi)^2} \left(\frac{-\partial f_0} {\partial \epsilon }\right) \frac{v_x^2({\bf k}) } {\tau^{\perp}} =
-\frac{ v_{F} k_{F} }  {4\pi \tau^{\perp}}
 \label{cond11c}
 \end{equation}
where $v_x({\bf k})=\partial \epsilon_{{\bf k}}^+ /\partial k_x$, $v_F$ is the Fermi velocity, and 
 \begin{equation}
 \begin{array}{l}
 1/\tau^{tr}=\int \frac{d^2{\bf k^{\prime}}}{(2\pi)^2}
  \, \omega_{{\bf k,k'}} \left(1-cos (\phi-\phi')\right)\\
 \\
 1/\tau^{\perp}= \int \frac{d^2{\bf k^{\prime}}}{(2\pi)^2}  \, \omega_{{\bf k,k'}} sin (\phi-\phi').
 \end{array}
 \label{taus}
 \end{equation}
Since the scattering rate $\omega_{{\bf k,k'}}$ is usually only weakly chiral ($\tau^{tr} \ll \tau^{\perp}$) 
$\omega_{{\bf k,k'}}$ can be estimated from time-dependent perturbation theory \cite{{Luttinger},{Leroux}}.
The lowest order symmetric scattering rate is given by the Golden rule expression,
while the lowest order antisymmetric contribution appears at third order (see, for example, Eqs. (2.7) and (3.11) in Ref.\onlinecite{Luttinger}).
\begin{eqnarray}
\omega^{(3a)}_ {{\bf k,k'}} &=& -(2\pi)^2 \delta (\epsilon_{{\bf k}} -\epsilon_{{\bf k'}}) \int \frac{d^2{\bf k''}}{(2\pi)^2} 
\nonumber \\ 
&& {\rm Im} \langle V_{{\bf k,k'}} V_{{\bf k',k''}} V_{{\bf k'',k}}\rangle_{dis} \; \delta(\epsilon_{{\bf k}}.
 -\epsilon_{{\bf k''}}).
\label{okkpa}
\end{eqnarray}
This yields 
\begin{align}
\frac{1}{\tau^{tr}} &=\frac{(vk_F)^2+4\Delta^2}{4\tau^q((vk_F)^2+\Delta^2)}\\ 
\frac{1} {\tau^{\perp}} &= \frac{V_1^3} {(\tau^q)^2 nV_0^4}   \frac { \Delta (vk_F)^2} {8 [(vk_F)^2 +\Delta^2]^{3/2}},
\label{tautr1}
\end{align}
so that the skew scattering Hall conductivity contribution due to non-Gaussian disorder correlations is 
\begin{equation}
\begin{array}{l}
\sigma_{xy}^{sk}
= - \frac{ e^2 V_1^3 }{2\pi nV_0^4 } \frac{\ \Delta (v k_F)^4 }{  (4\Delta^2+(v k_F)^2)^2}.
\end{array}
\label{jt}
\end{equation}
The Hall conductivity contribution (\ref{jt}) is inversely proportional to the impurity concentration $n$, 
and therefore can in principle dominate in the clean limit.
Since the size of third disorder correlation moment in a particular sample is unlikely
to be reliably known
and can be exceedingly small, we 
expect that the relative importance of skew scattering will always have to be assessed 
experimentally. 

\noindent
{\em Application to Graphene}-- A finite charge Hall conductance requires broken time reversal 
symmetry.  In graphene the vanishing conductance results from cancellation between bands of 
opposite spin.  The Hall conductance we evaluate here could be measured in graphene if the Fermi 
levels in the two spin-$\uparrow$ and the two spin-$\downarrow$ bands differed.  It may be possible to 
generate spin polarization in graphene by optical orientation, by tunneling through ferromagnetic contacts, 
or by hyperfine coupling to polarized nuclei. 
We note that the $\hat z$-component of spin is expected to relax particularly slowly in graphene 
because of the planar character of the crystal and the $\pi$-character of the orbitals near the 
Fermi energy. The alternative of studying the physics we address here, 
by applying an external magnetic field, is not favorable since it
leads to an ordinary Hall effect in addition to the anomalous Hall effect.  
We estimate that the anomalous portion of the 
Hall conductance in an external field is smaller by a factor $\sim (\Delta/\hbar v k) \times
(1/v k_F \tau)^2$. 
When the chemical potentials of spin-up and spin-down electrons are different our 
Hall effect calculation for each band remains valid.  The total 
Hall current is therefore 
\begin{equation}
\sigma_{xy}^{AHE} = 2(\sigma_{xy}(\mu_{\uparrow})-\sigma_{xy}(\mu_{\downarrow}))
\label{ahe}
\end{equation}
where the coefficient $2$ reflects equal contributions from the $K$ and $K'$ valleys.

The Hall conductivity we evaluate appears in the spin-Hall response even in the absence of external 
magnetic fields.  To find the magnitude of the SHE one should remember that instead of charge $e$ 
we are interested in spin $\pm 1/2$ carried by electrons:
$\sigma^{SH}_{xy} = 4\sigma_{xy}/2e$.
Here the coefficient $4$ is due to the $4$ bands which contribute equally to the SHE. 
The spin-Hall effect could be measured by using 
ferromagnetic leads, in the extreme case measuring transport only in one 
spin subsystem.  For that case the charge Hall conductivity becomes $2\sigma_{xy}$.
We expect that the results we derive here are valid for\cite{comment}  $\epsilon_F \gtrsim \tau^{-1}$  
whereas the quantized spin Hall conductivity will be observable only if $\Delta \gtrsim \tau^{-1}$.
The value of $\tau^{-1}$ in current samples can be estimated roughly from measured mobilities\cite{graphene_exp}
which are roughly constant except for Fermi energies below $\sim 50 {\rm meV}$.  Associating the change in 
mobility at low carrier densities with disorder mixing between electron and hole bands implies a $\tau^{-1}$ value
of the same order.  The value of $\Delta$ is difficult to estimate accurately.  Based on the relevant 
potential energy and length scales Kane and Mele have estimated that $\Delta \sim 0.2 \,{\rm meV}$.  This 
is likely to be an overestimate since the splitting represents an average of spin-orbit interactions that 
vary in sign over the system.  We\cite{ourprb} have separately estimated on the basis of a tight-binding model
with atomic spin-orbit interactions and {\em ab initio} electronic structure calculations that 
$\Delta \sim 0.001 \,{\rm meV}$.  In any event, it appears clear that sample quality will need 
to improve substantially in order to realize the quantum spin Hall effect.    
As our calculation shows, however, the surprisingly large 
anomalous Hall conductivities that flow from the chiral graphene 
bands should still be measurable in the metallic regime.  
Skew-scattering contributions, if present, should be separable experimentally in gated samples 
on the basis of their distinct carrier density dependence. 

 {\it Acknowledgments}. The authors are grateful for useful discussions with 
L. Brey, P. Bruno, F.D.M. Haldane, C. Kane, Q. Niu, K. Nomura, D. Novikov, and S. Urazhdin.
This work was supported by the Welch Foundation, by DOE grant DE-FG03-02ER45958, and by
ONR-N000140610122.

\end{document}